\documentclass[aps,prd,twocolumn,nofootinbib,groupedaddress,amsfonts,floatfix]{revtex4}

\usepackage{graphicx}

\usepackage{setspace}

\begin{document}

\title{Continuous Gravitational Waves from Isolated Galactic Neutron Stars in the Advanced Detector Era}

\author{Leslie Wade${}^1$} 
\author{Xavier Siemens${}^1$}
\author{David L. Kaplan${}^1$}
\author{Benjamin Knispel${}^2$}
\author{Bruce Allen${}^2$}

\affiliation{${}^1$Department of Physics, University of Wisconsin - Milwaukee, P.O. Box 413, Milwaukee, Wisconsin 53201} 
\affiliation{${}^2$Max-Planck-Institut f\"{u}r Graviationsphysik (Albert-Einstein-Institut) and Leibniz Universit\"{a}t Hannover Callinstr. 38, 30167 Hannover, Germany}
\date{\today} 

\begin{abstract}
We consider a simulated population of isolated Galactic neutron stars. The rotational frequency of each neutron star evolves through a combination of electromagnetic and gravitational wave emission.  The magnetic field strength dictates the dipolar emission, and the ellipticity (a measure of a neutron star's deformation) dictates the gravitational wave emission.  Through both analytic and numerical means, we assess the detectability of the Galactic neutron star population and bound the magnetic field strength and ellipticity parameter space of Galactic neutron stars with or without a direct gravitational wave detection.  While our simulated population is primitive, this work establishes a framework by which future efforts can be conducted.
\end{abstract}

\pacs{}

\maketitle
 
 
\section{Background and Motivation}

Isolated neutron stars with non-axisymmetric deformations will continuously emit gravitational waves as they rotate \cite{MTW}.  Neutron stars also have strong dipolar magnetic fields that accelerate particles to relativistic energies \cite{PAH}.  Since these neutron stars can lose energy through the emission of electromagnetic and gravitational radiation, their rotational frequency slowly decreases over time.  The gravitational wave strain amplitude of rotating neutron stars has a strong dependence on the star's rotational frequency.  Though no gravitational wave detection has yet been reported, rapidly rotating isolated Galactic neutron stars are one of the most promising sources of continuous gravitational waves for ground based gravitational wave detectors.

Attempts to assess the detectability of gravitational waves from the Galactic neutron star population began with rough analytic estimates.  An argument presented in \cite{300yrs} by Thorne but credited to Blandford models the Galactic neutron star population as a uniformly populated two-dimensional disk of gravitars (neutron stars with gravitationally dominated frequency evolutions) all born at very high frequencies.  Using this simplistic model, he estimated a rough upper bound on the possible gravitational wave strain amplitude from a Galactic neutron star, $h_{\rm max}\sim10^{-25}$ \cite{300yrs}.  Blandford also surprisingly observed that the maximum gravitational wave amplitude is independent of the size of the star's deformation and rotational frequency.  His argument was revised in \cite{LIGO_NS} and again in \cite{BB}, which both found $h_{\rm max}\sim10^{-24}$.

This work was followed by more comprehensive attempts to assess the detectability of the Galactic neutron star population through population synthesis.  If the neutron star population can be accurately simulated, then the detectability of Galactic neutron stars can be determined.  In \cite{Palomba} Palomba was the first to assess the detectability of a simulated gravitar population by first and second generation gravitational wave detectors.  He incorporated realistic spatial, age, birth frequency, and kick velocity distributions, as well as a possible ellipticity distribution (though this is still largely unconstrained \cite{Palomba}).  He estimated the fraction of the neutron star population that would likely have to be gravitars in order for first or second generation detectors to make a direct gravitational wave detection.  Continued efforts by Knispel and Allen extended Blandford's argument to a simulated gravitar population similar to Palomba's  \cite{BB}.  They found that the maximum gravitational wave strain amplitude {\it does} have a strong dependence on the star's frequency and size of deformation when considering a more realistic neutron star population.  They set upper bounds, which depend on the population's ellipticity (a measure of a star's deformation) and rotational frequency, on the gravitational wave strain amplitude of the nearest source.

In this paper, we include electromagnetic emission as well as gravitational wave emission in the frequency evolution of neutron stars and investigate its effect on the population's detectability.  We use the simulated neutron star population in \cite{BB} and assign every neutron star a dipolar magnetic field as well as an ellipticity.  We then allow each star's frequency to evolve through the emission of both gravitational and electromagnetic radiation.  The paper is organized as follows.  In Section \ref{NS_Eq} we review the spin and strain evolution of neutron stars and revisit the upper bounds from the gravitar case.  In Section \ref{methods} we outline a Monte Carlo simulation used to assess the detectability of the Galactic neutron star population.  The results are then used to bound the magnetic field strength and ellipticity parameter space of isolated neutron stars with or without a direct gravitational wave detection.  In Section \ref{analytic} we present a rough analytic argument to which we compare our numerical results.  In Section \ref{conc} we summarize our main results.

\section{Spin and Strain Evolution of Neutron Stars}
\label{NS_Eq}

We use the simulated neutron star population from \cite{BB} to assess the detectability of gravitational waves emitted by isolated Galactic neutron stars.  It is important to note that, while the simulated population does not explicitly include recycled millisecond pulsars, it does not necessarily exclude them either.  Each star in our population is assigned a birth frequency, initial position, kick velocity, and age.  Stars are then independently evolved through the Galaxy's gravitational potential (see \cite{BB} for a more detailed description of the population).  Therefore an old star that has been recently recycled can just be thought of as a young star born with a high frequency.  We also consider a large enough range in magnetic field strength to accommodate recycled pulsars.  In this section, we review methods to find the spin frequency and gravitational wave strain amplitude of each star in our simulated population in order to assess its detectability.

If neutron stars only lose energy through gravitational and electromagnetic emission, their rotational frequency evolution is given by
\begin{equation}
\label{fdot}
\dot{\nu}=-\frac{512\pi^4}{5}\frac{GI}{c^5}\epsilon^2\nu^5-\frac{8\pi^2}{3}\frac{{R}^6}{c^3I}B^2\sin^2{\alpha}\mbox{ }\nu^3,
\end{equation}
in cgs units \cite{Esposito}\cite{Rea}\cite{BB}.  Here, $G$ is the gravitational constant, $c$ is the speed of light, $\nu$ is the star's rotational frequency, $R$ is the star's radius, $I=kMR^2$ is the moment of inertia about its rotational axis with $M$ being the star's mass and $k\approx2/5$ \cite{PAH}, $\epsilon=(I_1-I_2)/I$ is the ellipticity with $I_1$ and $I_2$ being the moments of inertia about the star's other two principle axes, $B$ is the dipolar magnetic field strength at the magnetic equator, and $\alpha$ is the angle between its magnetic pole and its axis of rotation\footnote{The second term in Equation \ref{fdot}, which is the frequency evolution due to electromagnetic emission, is derived from the simple model of a rotating dipole.  In \cite{Spitkovsky} Spitkovsky corrects this term such that a neutron star will still emit electromagnetically even if its magnetic pole and rotational axis are aligned.}.  We choose the canonical values of $R=10\mbox{ km}$ and $M=1.4 M_\odot$ for all neutron stars \cite{PAH}.  Because we only concern ourselves with order of magnitude estimates, we set $\sin^2\alpha=1$.  

Equation \ref{fdot} can be solved analytically for $\nu(t,\nu_0)$ in the limits where $\epsilon=0$ or $B=0$.  If $B=0$, a neutron star will only emit gravitationally.  Its frequency is
\begin{equation}
\label{f_gw}
\nu(t,\nu_0)=\left({\nu_0}^{-4}-4\gamma_{\rm gw}t\right)^{-1/4},
\end{equation}
where $t$ is the neutron star's age, $\nu_0=\nu(t=0)$ is the neutron star's birth frequency, and $\gamma_{\rm gw}=-512\pi^4 GIc^{-5}\epsilon^2/5$.  Equation \ref{f_gw} is a good approximation for the frequency of a gravitar.  The characteristic timescale (the approximate time for a neutron star with birth frequency $\nu_0\gg\nu$ to spin down to a frequency $\nu$) for gravitationally dominated emission is
\begin{equation}
\label{t_gw}
\tau_{\rm gw}=-\frac{\nu}{4\dot{\nu}}\approx290\mbox{ Myrs}\left(\frac{10^{-7}}{\epsilon}\right)^2\left(\frac{100\mbox{ Hz}}{\nu}\right)^4.
\end{equation}
If $\epsilon=0$, a neutron star will only emit electromagnetically.  Its frequency is
\begin{equation}
\label{f_dip}
\nu(t,\nu_0)=\left({\nu_0}^{-2}-2\gamma_{\rm dip}t\right)^{-1/2},
\end{equation}
where $\gamma_{\rm dip}=-8\pi^2R^6c^{-3}I^{-1}B^2\sin^2{\alpha}/3$.  Equation \ref{f_dip} is a good approximation for the frequency of a neutron star whose evolution is dominated by electromagnetic emission and whose characteristic timescale is
\begin{equation}
\label{t_dip}
\tau_{\rm dip}=-\frac{\nu}{2\dot{\nu}}\approx1,600\mbox{ yrs}\left(\frac{10^{12}\mbox{ G}}{B}\right)^2\left(\frac{100\mbox{ Hz}}{\nu}\right)^2.
\end{equation}
While $\epsilon$ is unknown, the dramatically different timescales between Equations \ref{t_gw} and \ref{t_dip} illustrate the difficultly in detecting isolated neutron stars: stars with reasonable magnetic fields spin down to low frequencies too rapidly to detect.  Therefore, gravitational wave detectors will likely only detect neutron stars with small magnetic fields or young neutron stars that have not yet spun down to low frequencies.

Not all neutron stars will have their frequency evolution dominated by either gravitational or electromagnetic emission.  For these stars, $\dot{\nu}$ cannot be integrated over time to solve for an analytic solution for $\nu(t,\nu_0|\epsilon,B)$.  However, \cite{BB} shows that Equation \ref{fdot} can instead be inverted to solve for $t(\nu,\nu_0|\epsilon,B)$.  Following \cite{BB}, we rewrite Equation \ref{fdot} as
\begin{eqnarray}
\label{fdot_simple}
\dot{\nu}&=&\gamma_{\rm gw}\nu^5+\gamma_{\rm dip}\nu^3\\
\label{fdot_rewritten}
&=&\gamma_{\rm dip}\left(\gamma \nu^5+\nu^3\right),
\end{eqnarray}
where $\gamma=\gamma_{\rm gw}/\gamma_{\rm dip}$.  Equation \ref{fdot_rewritten} can be solved for
\begin{equation}
\label{t}
t(\nu,\nu_0)=\frac{1}{2|\gamma_{\rm dip}|}\left[\frac{{\nu_0}^2-\nu^2}{{\nu_0}^2\nu^2}+\gamma\ln\left(\frac{\nu^2}{{\nu_0}^2}\left(\frac{1+{\nu_0}^2\gamma}{1+\nu^2\gamma}\right)\right)\right].
\end{equation}
If $\nu_0$, $t$, $\gamma_{\rm gw}$, and $\gamma_{\rm dip}$ are known, Equation \ref{t} can be solved numerically to find $\nu$ using root finding techniques \cite{gsl}. 


The strain amplitude of gravitational waves emitted by a neutron star at a radial distance $r$ away from a detector is given by
\begin{eqnarray}
\label{h}
h&=&16\pi^2\frac{GI}{c^4}\frac{\epsilon \nu^2}{r}\\
&\approx&4\times10^{-25}\left(\frac{\epsilon}{10^{-7}}\right)\left(\frac{\nu}{100\mbox{ Hz}}\right)^2\left(\frac{1\mbox{ kpc}}{r}\right),
\end{eqnarray}
assuming that the neutron star's sky location intersects a line normal to the plane of the detector arms and its axis of rotation is parallel to that line (optimal mutual orientation).  Since we only
concern ourselves with order of magnitude estimates, we assume optimal mutual orientation for all neutron stars \cite{BB}, which overestimates the detectable amplitude by about a factor of four on average.

For a population of neutron stars whose radial distance from Earth $r$, age $t$, birth frequency $\nu_0$, ellipticity $\epsilon$, and magnetic field strength $B$ are known, Equation \ref{f_gw}, \ref{f_dip}, or \ref{t} can be used to determine each star's spin frequency $\nu$.  Equation \ref{f_gw} is used when $\gamma_{\rm gw}\nu^5\gg\gamma_{\rm dip}\nu^3$, which we conservatively choose to be when $\gamma>40\mbox{ s$^2$}$; Equation \ref{f_dip} is used when $\gamma_{\rm gw}\nu^5\ll\gamma_{\rm dip}\nu^3$, which we conservatively choose to be when $\gamma<4\times10^{-9}\mbox{ s$^2$}$; Equation \ref{t} is used otherwise\footnote{To determine the two $\gamma$ cutoffs, we assume that one term will dominate over the other if it is at least three orders of magnitude greater than the other.  Equation \ref{f_gw} can be used when $\gamma\gg1/\nu^2$.  Therefore, we choose $\gamma>10^3/\nu^2=40\mbox{ s}^2$ for $\nu=5\mbox{ Hz}$.  Equation \ref{f_dip} can be used when $\gamma\ll1/\nu^2$.  Therefore, we choose $\gamma<10^{-3}/\nu^2=4\times10^{-9}\mbox{ s}^2$ for $\nu=500\mbox{ Hz}$.}.  Equation \ref{h} can further be used to determine each star's gravitational wave strain amplitude $h$ as measured in our detector.  We compare each star's frequency and strain amplitude to a scaled gravitational wave detector's noise curve in order to assess the detectability of the neutron star population.  We explain how we derive the scaling factor in Section \ref{methods}.

\begin{figure}[t]
\includegraphics[width=3.4in]{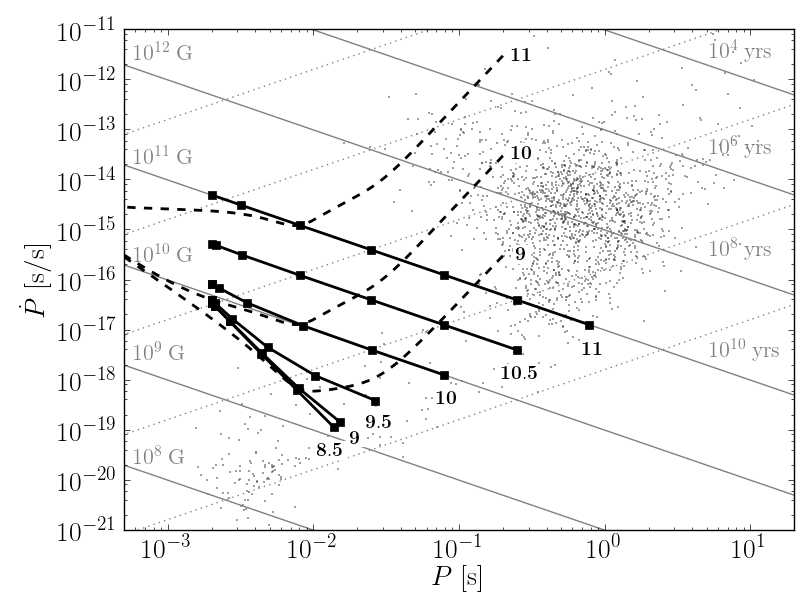}
\caption{\label{P_Pdot}{\footnotesize This figure shows the period evolution of neutron stars.  The dots represent observed pulsars from the ATNF catalog \cite{ATNF}.  The thin, negatively sloped solid contours are lines of constant magnetic field strength (labels on the left), and the thin, positively sloped dotted contours are lines of constant characteristic age (labels on the right) assuming only electromagnetic emission.  The thick, solid lines with square ticks track the period evolution of a neutron star that emits both electromagnetic and gravitational radiation.  These lines, which are labelled by the logarithm of the star's magnetic field in units of Gauss, correspond to a neutron star with $\epsilon=10^{-7}$ (all lines) and $B=10^{8.5}-10^{11}$ with steps of $1/2$ dex.  The square ticks represent logarithmic steps in age.  The first tick labels $t=0$, and the subsequent ticks range from $t=10^4-10^9\mbox{ yrs}$.  The thick, dashed lines, which are labelled by the logarithm of the star's magnetic field in units of Gauss, are characteristic aLIGO sensitivity curves for neutron stars with $\epsilon=10^{-7}$ located $100\mbox{ pc}$ away from Earth.  Neutron stars below their associated aLIGO sensitivity curve are undetectable.  Neutron stars with large magnetic fields spin down to low frequencies (high periods) much faster than stars with small magnetic fields; consequently, they spend less of their lives emitting gravitational waves with frequencies that aLIGO is most likely to detect.}}
\end{figure}

While Equation \ref{h} for the gravitational wave strain amplitude $h$ does not depend explicitly on the magnetic field, $B$ {\it does} help to determine $\nu$ through Equation \ref{fdot}.  There are two related effects.  First, Figure \ref{P_Pdot} shows that, all other things being equal, neutron stars with large magnetic fields will spin down to low frequencies (high periods) much faster than neutron stars with small magnetic fields.  Consequently, large magnetic fields will result in smaller and smaller gravitational wave amplitudes over time.  Second, since gravitational wave detectors are sensitive to finite frequency ranges, neutron stars with large magnetic fields will rapidly spin through a detector's sensitive frequencies, which makes them less likely to be detected.  Therefore, neutron stars with small magnetic fields are more likely to be detected than neutron stars with large magnetic fields.  In this way, assuming we know the population's ellipticity, we can place lower bounds on the magnetic field of neutron stars in the absence of a gravitational wave detection.

\begin{figure}[t]
\includegraphics[width=3.4in]{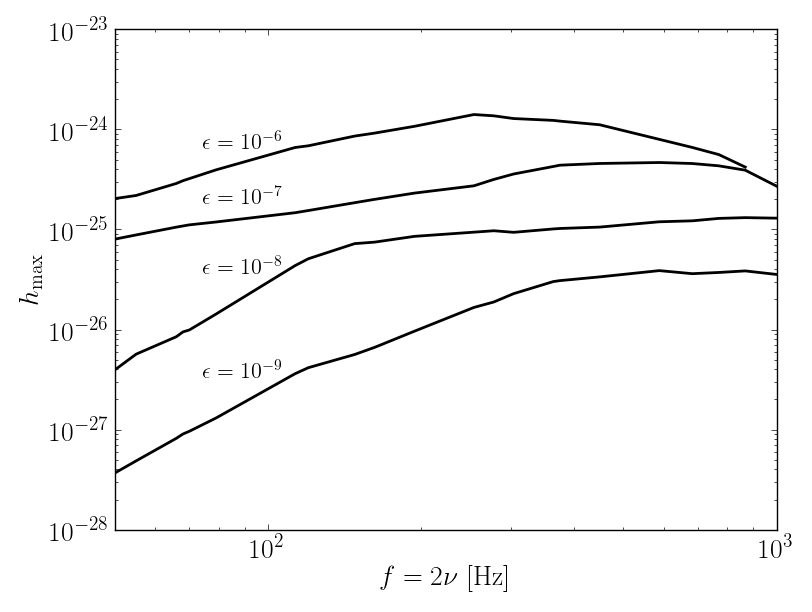}
\caption{\label{hmax_vs_f}{\footnotesize  We plot the maximum gravitational wave strain amplitudes $h_{\rm max}$ as a function of gravitational wave frequency $f=2\nu$ of a population of gravitars (neutron stars with $B=0$) with $\epsilon=10^{-9}$, $10^{-8}$, $10^{-7}$, and $10^{-6}$.  A single point $(f,h_{\rm max})$ corresponds to the population's maximum gravitational wave amplitude $h_{\rm max}$ measured in the frequency band $[f,ef]$.  We used the gamma initial radial distribution from \cite{BB} to simulate the neutron star population.}}
\end{figure}

We can gain intuition into the detectability of Galactic neutron stars by setting $B=0$.  This places an upper bound on $h$ for fixed $\epsilon$ values.  In Figure \ref{hmax_vs_f}, we plotted the maximum gravitational wave strain amplitude $h_{\rm max}$ versus gravitational wave frequency $f=2\nu$ of the simulated neutron star population presented in \cite{BB} with $B=0$ and $\epsilon=10^{-9}$, $10^{-8}$, $10^{-7}$, and $10^{-6}$.  A single point $(f,h_{\rm max})$ corresponds to the population's maximum gravitational wave amplitude $h_{\rm max}$ measured in the frequency band $[f,ef]$ where $e$ is Euler's number.\footnote{To find $h_{\rm max}$, we considered 200 logarithmically spaced overlapping frequency bands and constructed histograms for the strain amplitude from the neutron stars in each band.  We then solved for $h_{\rm max}$ using a linear fit in log$_{10}$-space to the tail (largest $h$ values) of each histogram.  We used this method to minimize statistical fluctuations.}

Our numerical result in Figure \ref{hmax_vs_f} is consistent with the result in \cite{BB}, which was derived using a semi-analytical integration technique. Considering a distribution in frequency and a three-dimensional spatial distribution results in a clear dependence of $h_{\rm max}$ on both frequency and ellipticity \cite{BB}.  The effect of the frequency distribution manifests itself in the overall shape of the four curves in Figure \ref{hmax_vs_f}.  Since stars with large ellipticities spin down much faster than stars with small ellipticities (Equation \ref{t_gw}), a neutron star population with large ellipticities more densely populates low frequency bands than a neutron star population with small ellipticities.  Therefore, while each curve in Figure \ref{hmax_vs_f} has a similar shape, the large ellipticity curves are shifted to the left relative to the small ellipticity curves.  The subtle kink in the $\epsilon=10^{-9}$ and $10^{-8}$ curves between the nearly flat, high frequency region and the more positively sloped, low frequency region corresponds to a kink in the population's frequency distribution, which is described in \cite{BB}.\footnote{Since we consider a continuous distribution in birth frequency, and a single star cannot be older than the Galaxy, neutron stars with high birth frequencies will not have existed long enough to have spun down past a certain frequency.  Neutron stars will accumulate near this frequency causing a kink in the population's frequency distribution, as seen in \cite{BB}.}  The $\epsilon=10^{-7}$ and $10^{-6}$ curves also exhibit the same behavior but at smaller frequencies than those plotted.  If we had considered a two-dimensional spatial distribution, $h_{\rm max}$ would have been independent of the ellipticity in the region to the right of the kink.  Here, the frequency distribution is in a nearly steady state.  Considering a three-dimensional spatial distribution breaks this degeneracy between $h_{\rm max}$ and the population's ellipticity.  Note that the gravitational wave strain amplitude will decrease when magnetic fields are considered.

\section{Neutron Star Detectability and Constraints}
\label{methods}

To assess the detectability of the Galactic neutron star population, we use the methods outlined in Section \ref{NS_Eq} to find the spin frequency $\nu$ and gravitational wave strain amplitude $h$ of every neutron star in our simulated population.  The population used in our analysis is described in detail in \cite{BB}.  Although \cite{BB} presents three different initial radial distributions, we choose to present our results using just the ``gamma initial radial distribution''.  We expect minimal deviation in our results if we were to consider the other two distributions presented in \cite{BB}.  The code that creates our simulated population provides the final position, which is easily turned into a distance from Earth $r$, and age $t$ of each neutron star.  Since a star's magnetic field strength $B$ and ellipticity $\epsilon$ dictate its frequency evolution, we must also choose what values of each to assign to the stars in our population.  We fix $\epsilon$ and $B$ to a single value so that every neutron star in our population has the same $\epsilon$--$B$ combination.  While this approach will not mimic a realistic neutron star population, it is an important first step that provides valuable intuition for considering a more realistic population in the future.  Lastly, we assign each star a birth frequency $\nu_0=1/P_0$, where $P_0$ is randomly drawn from the lognormal birth period distribution in \cite{BB}:
$$
\rho_{\rm P_0}(P_0)=\frac{1}{\sqrt{2\pi}\sigma P_0}\exp\left[-\frac{1}{2\sigma^2}\left(\ln{P_0}-\ln\bar{P_0}\right)^2\right].
$$
Here, $P_0>0.5\mbox{ ms}$ is the birth period, $\bar{P_0}=5\mbox{ ms}$ is the mean, and $\sigma=0.69$ is the standard deviation.  Given $r$, $t$, $\nu_0$, $\epsilon$, and $B$, we can use the methods outlined in Section \ref{NS_Eq} to find $\nu(t,\nu_0|\epsilon,B)$ and $h(r,\nu|\epsilon)$ for every neutron star in our simulated population.

\begin{figure}[t]
\includegraphics[width=3.4in]{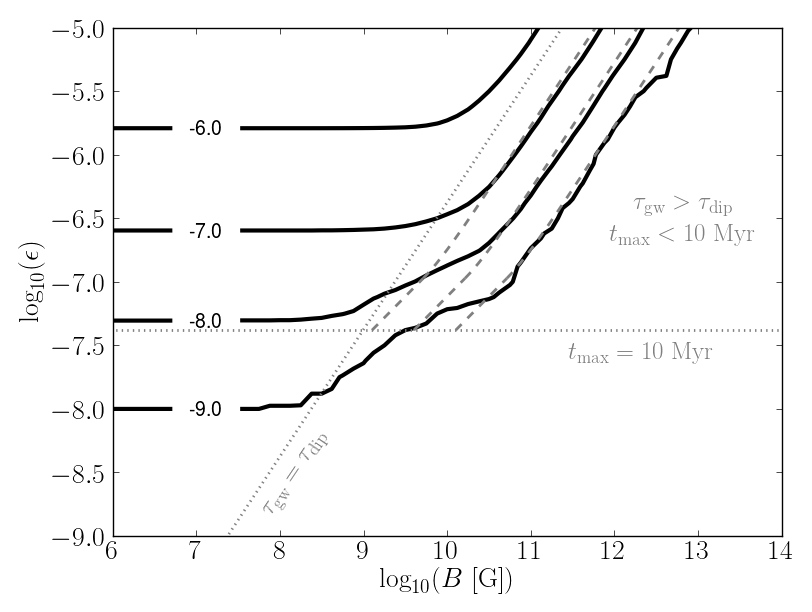}
\caption{\label{frac_contour}{\footnotesize Contours of $\log_{10}n$ (described in Section \ref{methods}) with respect to $(\log_{10}B,\log_{10}\epsilon)$, which illustrate the expected detectability of the neutron star population with various $\epsilon$ and $B$ combinations.  Our analysis was performed on populations with logarithmic spacings of 1/8 dex in $\epsilon$ and $B$.  The dashed lines are the analytic approximations for $\log_{10}n$ described in Section \ref{analytic}.  We plot three results: $\log_{10}n=-7, -8, -9$, respectively from left to right.  The dotted lines show the boundaries that separate where the analytic argument's assumptions are valid from where they are not (Section \ref{analytic}).  They only hold for detectable neutron stars that are young ($t_{\rm max}\lesssim10\mbox{ Myrs}$), found above the horizontal dotted line, and dominated by electromagnetic emission ($\tau_{\rm dip}\lesssim\tau_{\rm gw}$), found to the right of the positively sloped dotted line.}}
\end{figure}

Once $\nu$ and $h$ are found for every neutron star, we can determine whether or not we expect a gravitational wave detector to detect our population.  For simplicity, we only consider detection by a single Advanced Laser Interferometer Gravitational-wave Observatory (aLIGO) detector.  We use the aLIGO noise curve for a single detector from \cite{LIGOnoise}, which is the expected sensitivity of aLIGO as a function of gravitational wave frequency.  To estimate the strain, we assume that we have a year of aLIGO data, and that the data is analyzed coherently in short 72 hr stretches, with the short stretches combined incoherently.  This assumes the LIGO Scientific Collaboration will be doing similar searches to the ones currently done by Einstein@Home \cite{E@H} in the aLIGO era.  An overall trials factor of $100$ is applied, which is considered a conservative estimate.  We then compare each neutron star to the estimated noise curve to determine the number of neutron stars in our population that aLIGO will be able to detect.  We assume that a neutron star will be detected if its strain is above the aLIGO noise curve.  To assess the detectability of the neutron star population, we construct the fraction
\begin{equation}
\label{n}
n=\frac{N_{\rm det}}{N_{\rm sim}},
\end{equation}
where $N_{\rm sim}$ is the number of stars in the simulated population, and $N_{\rm det}$ is the number of stars aLIGO can detect from this population.  To reduce statistical fluctuations, we simulate many more neutron stars than are actually expected to be in our Galaxy.  Multiplying this fraction $n$ by the number of neutron stars in our Galaxy $N_{\rm gal}$ gives the number of detectable neutron stars in our Galaxy.  If $n\cdot N_{\rm gal}$ is greater than or equal to one, the population will likely be detectable; if it is less than one, the population will likely be undetectable.  In Figure \ref{frac_contour}, we plot contours of $\log_{10}n$ with respect to ($\log_{10}B$, $\log_{10}\epsilon$), illustrating the expected detectability of the neutron star population with various $\epsilon$ and $B$ combinations.

We can further use our results (Figure \ref{frac_contour}) to place bounds on the $\epsilon$--$B$ parameter space of the Galactic neutron star population.  In Figure \ref{frac_contour}, the contour corresponding to $n\cdot N_{\rm gal}=1$ is the boundary above which the $\epsilon$--$B$ parameter space is disallowed, assuming (pessimistically) no aLIGO detection of continuous gravitational waves associated with a Galactic neutron star.  In this way, Figure \ref{frac_contour} sets lower bounds on $B$ for fixed $\epsilon$ values (or upper bounds on $\epsilon$ for fixed $B$ values) if aLIGO does not make an isolated neutron star detection.  For instance if $N_{\rm gal}\sim10^9$ \cite{Palomba}, and we assume neutron stars have a typical ellipticity of $\epsilon\sim10^{-7}$ \cite{Palomba}, Figure \ref{frac_contour} shows that the minimum magnetic field strength of Galactic neutron stars is $B\gtrsim10^{11}\mbox{ G}$ in the absence of an aLIGO detection.  Conversely if $N_{\rm gal}\sim10^9$, and we assume neutron stars have a typical magnetic field strength of $B\sim10^{11}\mbox{ G}$, Figure \ref{frac_contour} shows that the population's maximum ellipticity is $\epsilon\lesssim10^{-7}$ in the absence of an aLIGO detection.  This argument also applies if aLIGO {\it does} make isolated neutron star detections.  If $N_{\rm gal}\sim10^9$, and we assume neutron stars have $\epsilon\sim10^{-7}$, then the minimum magnetic field strength of Galactic neutron stars is $B\gtrsim10^{10}\mbox{ G}$ if aLIGO detects $10$ neutron stars.  Conversely if $N_{\rm gal}\sim10^9$, and we assume neutron stars have $B\sim10^{10}\mbox{ G}$, then the population's maximum ellipticity is $\epsilon\lesssim10^{-7}$ if aLIGO detects $10$ neutron stars.

\section{Analytic Results}
\label{analytic}

In all previous sections, we used numerical methods to assess the detectability of Galactic neutron stars and place constraints on the properties of the Galactic neutron star population.  In this section, we present an analytical approach to setting bounds on the $\epsilon$--$B$ parameter space of the Galactic neutron star population.  Blandford's analytic argument considers neutron stars that emit only gravitationally.  Our analytic argument, while still simplistic, applies to stars dominated by electromagnetic emission.

\begin{figure}[t]
\includegraphics[width=3.4in]{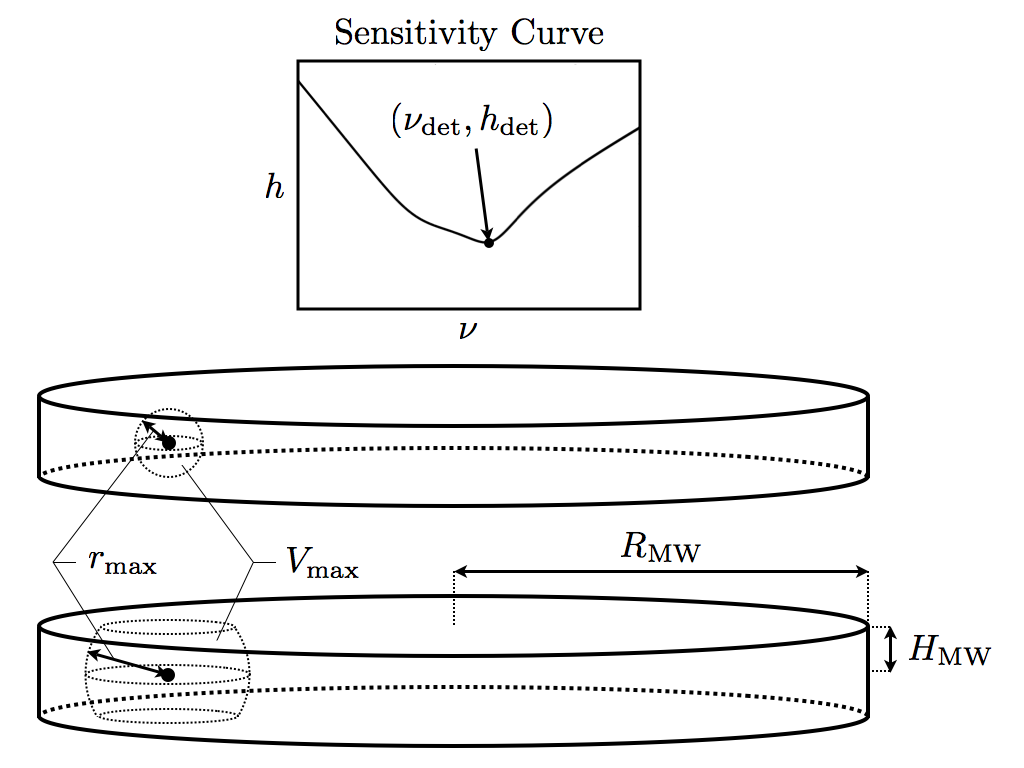}
\caption{\label{an_diagram}{\footnotesize The top plot is a typical sensitivity curve of a ground-based gravitational wave detector.  We assume that isolated neutron stars with $h(\nu_{\rm det})>h_{\rm det}$ will be detected.  Below are diagrams of the Milky Way disk.  The black dot is Earth's location within the Milky Way disk.  The two volumes $V_{\rm max}$ outlined by dotted lines are the maximum volumes within which detectable neutron stars can be contained.  $V_{\rm max}$ will be the volume of a sphere if $r_{\rm max}<H_{\rm MW}$ (top diagram), and $V_{\rm max}$ will be the volume of a spherical segment if $r_{\rm max}>H_{\rm MW}$ (bottom diagram).  See Section \ref{analytic}.}}
\end{figure}

We first use the aLIGO sensitivity curve described in Section \ref{methods} to constrain the volume around Earth in which detectable neutron stars must be contained.  If a neutron star is detected, it will tend to be at or near the detector's most sensitive frequency, which we call $\nu_{\rm det}$.  For simplicity, we assume that a neutron star must have $\nu\approx\nu_{\rm det}$ to be detected.  Therefore, a neutron star will be detected by a ground-based gravitational wave detector if $h(\nu_{\rm det})>h_{\rm det}$, where $h_{\rm det}$ is the value of the strain amplitude for which the detector is most sensitive (see Figure \ref{an_diagram}).  This detectability condition, along with Equation \ref{h}, translates into a constraint on the distance from Earth of detectable neutron stars.  The maximum distance $r_{\rm max}$ at which a neutron star with $\nu=\nu_{\rm det}$ could be detected is: 
\begin{equation}
r_{\rm max}=16\pi^2\frac{GI}{c^4}\frac{\epsilon \nu_{\rm det}^2}{h_{\rm det}}.
\end{equation}

The detectability condition also translates into a constraint on the volume that encloses detectable neutron stars.  First, we assume that neutron stars are born uniformly throughout the Galactic stellar disk at a constant rate ${\cal N}$, which is the number of births per unit time.  The volume of the Milky Way, which we approximate to be a disk, is roughly
\begin{equation}
V_{\rm MW}=\pi R_{\rm MW}^2(2H_{\rm MW}),
\end{equation}
where $R_{\rm MW}$ is the radius of the Galactic disk, and $H_{\rm MW}$ is half its height (Figure \ref{an_diagram}).  The volume contained in $r_{\rm max}$ will be a sphere for $r_{\rm max}<H_{\rm MW}$.  However, for $r_{\rm max}>H_{\rm MW}$, the volume contained within $r_{\rm max}$ will be a sphere with its top and bottom caps truncated by the top and bottom surfaces of the Milky Way disk, as illustrated in Figure \ref{an_diagram}.  This shape is called a spherical segment.  Therefore, the maximum volume $V_{\rm max}$ in which neutron stars with $\nu=\nu_{\rm det}$ could be detected is:
\begin{equation}
V_{\rm max}=\left\{
\begin{array}{l l}
\frac{4}{3}\pi r_{\rm max}^3 & r_{\rm max}<H_{\rm MW}\\
\\
\frac{2}{3}\pi H_{\rm MW} \left(3r_{\rm max}^2-H_{\rm MW}^2\right) & r_{\rm max}>H_{\rm MW}\\
\end{array}
\right..
\end{equation}

From the constraint on the volume that encloses detectable neutron stars of frequency $\nu=\nu_{\rm det}$, we can find the minimum allowed magnetic field strength in the absence of an isolated neutron star detection.  Remembering our constant birth rate assumption, the average time $t_{\rm max}$ between neutron star births into the volume $V_{\rm max}$ is
\begin{equation}
t_{\rm max}={\cal N}_{\rm max}^{-1}=\frac{V_{\rm MW}}{V_{\rm max}}{\cal N}^{-1},
\end{equation}
assuming a uniform spatial distribution.  In order to ensure a neutron star detection, at least one star within the volume $V_{\rm max}$ must have $\nu>\nu_{\rm det}$ at all times.  This will be the case if $\nu(t_{\rm max})>\nu_{\rm det}$, because a neutron star spinning down below $\nu_{\rm det}$ will always be accompanied by a new star being born into the volume $V_{\rm max}$.  Likewise, we also assume that, on average, when a detectable neutron star escapes $V_{\rm max}$ due to its motion in the Galaxy, another detectable neutron star will enter $V_{\rm max}$. Assuming that a neutron star's frequency evolution is dominated by dipolar emission, we solve for the minimum magnetic field strength $B_{\rm min}$ below which a neutron star detection is {\it not} guaranteed by substituting $t_{\rm max}$ into Equation \ref{f_dip} and solving for $B$:
\begin{equation}
\label{B_min}
B_{\rm min}(\epsilon,h_{\rm det},\nu_{\rm det})=\left\{
\begin{array}{l l}
B_{\rm min}^{\rm sphere} & r_{\rm max}<H_{\rm MW}\\
\\
B_{\rm min}^{\rm sph. seg.} & r_{\rm max}>H_{\rm MW}\\
\end{array}
\right.,
\end{equation}
where
\begin{eqnarray}
B_{\rm min}^{\rm sphere}&=&\frac{32 \pi^2 \nu_{\rm det}^2 I^2\epsilon^{3/2}}{\nu_0 R^3 h_{\rm det}^{3/2}}\sqrt{\frac{\pi G^3{\cal N}}{c^9V_{\rm MW}} \left(\nu_0^2-\nu_{\rm det}^2\right)}\nonumber \\
B_{\rm min}^{\rm sph. seg.}&=&\frac{cH_{\rm MW}}{2R^3\nu_0\nu_{\rm det}}\sqrt{\frac{c I {\cal N}H_{\rm MW}}{2\pi V_{\rm MW}}\left(\nu_0^2-\nu_{\rm det}^2\right)}\nonumber \\
&&\times\left[\frac{768\pi^4G^2I^2\epsilon^2\nu_{\rm det}^4}{c^8h_{\rm det}^2H_{\rm MW}^2}-1\right]^{1/2} .\nonumber
\end{eqnarray}
Neutron stars with $B>B_{\rm min}$ in $V_{\rm max}$ will spin down to $\nu<\nu_{\rm det}$ before another star is born into $V_{\rm max}$.  Therefore, in the absence of an aLIGO detection, $B=B_{\rm min}$ is the minimum possible magnetic field strength of Galactic neutron stars with fixed ellipticity values, since $B<B_{\rm min}$ ensures a detection.

Our argument is easily extended to the case of $N_{\rm s}$ neutron star detections.  To do this, we solve for when $\nu(N_{\rm s}t_{\rm max})>\nu_{\rm det}$.  The result adds a factor of $N_{\rm s}^{-1/2}$ in front of Equation \ref{B_min}.

We have made several assumptions in setting up our analytical argument.  It is important to emphasize two of our argument's most crucial assumptions in order to clearly outline the physical systems for which our argument holds.  The first crucial assumption is that the spatial distribution of neutron stars in the Milky Way is a uniform cylinder.  Neutron stars will diffuse out of the Galactic disk due to Galactic acceleration and their kick velocities.  The timescale for this process is found by dividing the average kick velocity in \cite{BB} by the gravitational acceleration (found by dividing the gravitational potential in \cite{BB} by the length scale).  Therefore, our argument holds when $t_{\rm max}\lesssim10\mbox{ Myrs}$.  The second crucial assumption is that the frequency evolution of neutron stars is dominated by dipolar emission.  Therefore, for $\nu_0\gg\nu$, our argument holds when $\tau_{\rm dip}\lesssim\tau_{\rm gw}$.

In Figure \ref{frac_contour}, we have plotted the relationship in Equation \ref{B_min} (with the factor of $N_{\rm s}^{-1/2}$ in front) as dashed lines on top of our numerical results.  We use $\nu_{\rm det}\approx100\mbox{ Hz}$ and $h_{\rm det}\approx6.0\times 10^{-26}$, which approximately corresponds to aLIGO's most sensitivity strain and associated frequency, and $R_{\rm MW}\approx15\mbox{ kpc}$, $H_{\rm MW}\approx75\mbox{ pc}$, and $\nu_0\approx850\mbox{ Hz}$, where $R_{\rm MW}$, $H_{\rm MW}$, and $\nu_0$ are estimated averages of the spatial and period distributions in \cite{BB} found by reducing the maximum values by a factor of $e^{-1}$.  We also use ${\cal N}\approx0.02\mbox{ years}^{-1}$ \cite{Diehl}.  Our numerical results should roughly follow these dashed lines, which correspond to $n=10^{-7}, 10^{-8}, 10^{-9}$, respectively from left to right.  The analytic results only hold for detectable neutron stars that are young ($t_{\rm max}\lesssim10\mbox{ Myrs}$), which corresponds to the region above the horizontal dotted line, and dominated by electromagnetic emission ($\tau_{\rm dip}\lesssim\tau_{\rm gw}$), which corresponds to the region to the right of the positively sloped dotted line.  There is good agreement between our numerical and analytic results, except in the transition region near the dotted boundaries where the analytic assumptions start to lose their validity.  While the rough numerical values chosen for the parameters in our analytic argument can change the overall normalization of the analytic curves, the shape of the curves closely match the shape of the numerical contours.

\begin{figure}[t]
\includegraphics[width=3.4in]{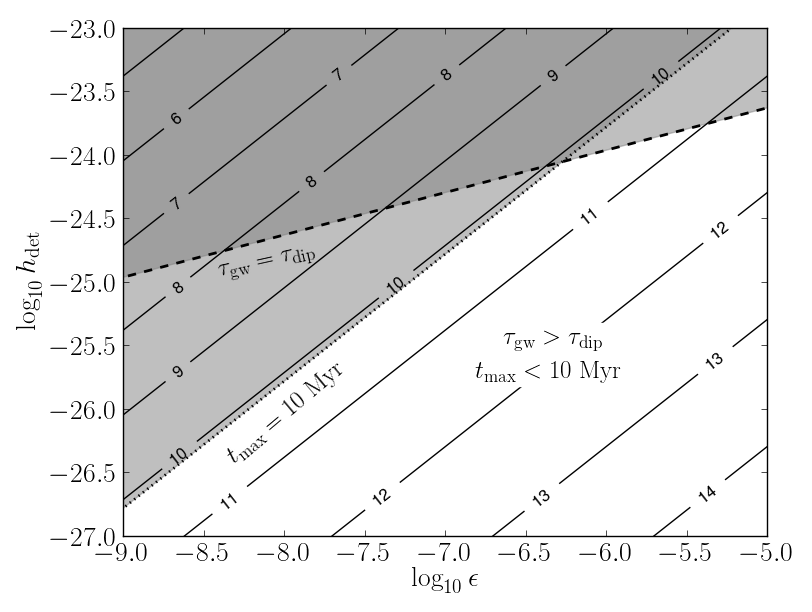}
\caption{\label{hmax_epsilon}{\footnotesize Extends our analytic argument to any gravitational wave detector.  Equation \ref{B_min} is plotted as solid contours of $\log_{10}B$ with respect to $(\log_{10}h_{\rm det},\log_{10}\epsilon)$, where $B$ is in units of Gauss.  The dashed and dotted lines show the boundaries that separate where our analytic argument's assumptions are valid from where they are not.  Our argument does not hold in the gray, shaded regions; our argument {\it does} hold for neutron stars that are young ($t_{\rm max}\lesssim10\mbox{ Myrs}$), found below the dotted line ($t_{\rm max}=10\mbox{ Myrs}$), and dominated by electromagnetic emission ($\tau_{\rm dip}\lesssim\tau_{\rm gw}$), found below the dashed line ($\tau_{\rm dip}=\tau_{\rm gw}$).}}
\end{figure}

While we only consider detection by aLIGO in our numerical analysis, our analytical approach easily extends to any gravitational wave detector.  Notice that in Equation \ref{B_min} $B_{\rm min}$ is a function of $\epsilon$, $h_{\rm det}$, and $\nu_{\rm det}$.  Therefore, we fix $\nu_{\rm det}$ and plot contours of $\log_{10}B_{\rm min}$ with respect to $(\log_{10}\epsilon,\log_{10}h_{\rm det})$ in Figure \ref{hmax_epsilon} to illustrate how our argument extends to other detectors.

It also seems natural to extend our argument to the gravitar case, in which the frequency evolution of neutron stars is dominated by gravitational emission, by solving Equation \ref{f_gw} for $B_{\rm min}$ under the assumption that $\tau_{\rm gw}\lesssim\tau_{\rm dip}$.  However, detectable gravitars can be older than $10\mbox{ Myrs}$, thus violating our assumption that $t_{\rm max}\lesssim10\mbox{ Myrs}$.  Therefore, these methods cannot be applied to the gravitar case.

\section{Conclusion}
\label{conc}

We have used the methods described in Section \ref{NS_Eq} to find the gravitational wave amplitude and spin frequency of every neutron star in the simulated population described in \cite{BB}.  This involved allowing for both electromagnetic and gravitational emission in a neutron star's frequency evolution (Equation \ref{fdot}).  We then solved for each neutron star's frequency using either Equation \ref{f_gw}, \ref{f_dip}, or \ref{t} and each neutron star's gravitational wave strain amplitude using Equation \ref{h}.  We used the simulated population to assess the detectability of and set bounds on the $\epsilon$--$B$ parameter space of the Galactic neutron star population.  Our results are summarized in Figure \ref{frac_contour}.  Assuming that the Galactic neutron star population consists of $N_{\rm gal}\sim10^9$ stars, and assuming aLIGO does not make a neutron star detection, the contour $\log_{10}n=-9$ in Figure \ref{frac_contour} separates the allowed $\epsilon$--$B$ parameter space (below the contour) from the disallowed $\epsilon$--$B$ parameter space (above the contour).  In other words, assuming we know the magnetic field strength of the neutron star population, we can place upper bounds on the population's ellipticity;  or, assuming we know the ellipticity of the neutron star population, we can place lower bounds on the population's magnetic field strength.

In this paper, we have only considered the simple (and unrealistic) case in which all neutron stars have the same magnetic field strength and ellipticity.  However, we have demonstrated that both a gravitational wave detection or the lack of a gravitational wave detection can be used to constrain some of the properties of the Galactic neutron star population.  To make confident quantitative statements regarding the properties of the Galactic neutron star population, we must construct a more realistic population.  Moving forward, we plan to incorporate magnetic field and ellipticity distributions and evolutions into our analysis to more closely mimic the Galactic neutron star population \cite{Palomba}\cite{2009A&A...496..207P}\cite{2010MNRAS.401.2675P}.

\section{Acknowledgments}

L.W. would like to thank Patrick Brady, Jolien Creigthon, and Madeline Wade for helpful discussions and Adam Mercer, Chris Pankow, and Greg Skelton for diligent technical assistance.  X. S. would like to thank Curt Cutler for useful discussions.  We would also like to thank the anonymous referee who provided several helpful comments and suggestions.  This work was partially funded by the NSF through CAREER award number 0955929 and award number 0970074, and the Wisconsin Space Grant Consortium. 

\bibliographystyle{unsrt} 
\bibliography{biblio}

\end{document}